\documentclass[11pt,draftclsnofoot,journal,onecolumn]{IEEEtran}

\usepackage[dvips]{graphicx}
\usepackage{subfigure}
\usepackage{amsmath}
\usepackage{amssymb}
\usepackage{eurosym}
\usepackage{cases}
\usepackage{color}
\usepackage{url}
\usepackage{cite}
\usepackage[ruled]{algorithm2e}
\SetAlFnt{\footnotesize}
\SetAlCapFnt{\footnotesize}
\SetAlCapNameFnt{\footnotesize}
\usepackage{algorithmic}
\usepackage{subeqnarray}
\usepackage{cases}
\usepackage{flushend}

\newtheorem{proposition}{Proposition}

\renewcommand{\IEEEQED}{\IEEEQEDopen}

\begin{document}

\title{Planning of Cellular Networks Enhanced by Energy Harvesting}
\author{Meng Zheng, Przemys{\l}aw Pawe{\l}czak, S{\l}awomir Sta\'{n}czak,~and Haibin Yu
\thanks{Meng Zheng and Hiabin Yu are with the Laboratory of Industrial Control Network and System, Shenyang Institute of Automation, Chinese Academy of Sciences, 114 Nanta Street, Shenyang, 110016 China (email: \{zhengmeng\_6, yhb\}@sia.cn). Meng Zheng was also with the Fraunhofer Institute for Telecommunication, Heinrich-Hertz-Institute, Berlin, Germany.}
\thanks{Przemys{\l}aw Pawe{\l}czak is with the Delft University of Technology, Mekelweg 4, 2628 CD Delft, The Netherlands (email: p.pawelczak@tudelft.nl).}
\thanks{S{\l}awomir Sta\'{n}czak is with the Fraunhofer Institute for Telecommunication, Heinrich-Hertz-Institute, Einsteinufer 37, 10587 Berlin, Germany (email: slawomir.stanczak@hhi.fraunhofer.de).}
\thanks{This work has been supported in part by the German Federal Ministry of Economics and Technology under grant 01ME11024. The first author was partially supported by the Natural Science Foundation of China under grant 61174026 and 61233007. Part of this paper appeared as a poster in~\cite{pawelczak_dyspan_2012}.}
\thanks{\copyright 2013 IEEE. Personal use of this material is permitted. Permission from IEEE must be 
obtained for all other uses, in any current or future media, including  reprinting/republishing this material for advertising or promotional purposes, creating new  collective works, for resale or redistribution to servers or lists, or reuse of any copyrighted  component of this work in other works.}
}

\maketitle

\begin{abstract}
We pose a novel cellular network planning problem, considering the use of renewable energy sources and a fundamentally new concept of energy balancing, and propose a novel algorithm to solve it. In terms of the network capital and operational expenditure, we conclude that savings can be made by enriching cellular infrastructure with energy harvesting sources, in comparison to traditional deployment methods.
\end{abstract}

\IEEEpeerreviewmaketitle
\vspace{-0.5cm}

\section{Introduction}
\label{sec:Int}

\IEEEPARstart{A}{lthough} equipping base stations (BSs) with Renewable Energy Sources (RESs), such as solar panels and wind turbines (either to support or replace traditional electric grid connection with energy harvesting~\cite{ng_submitted_2013}), has been proved technically feasible, the research on RES-powered cellular networks is still in its infancy. One of the fundamental questions to be answered is how to plan the topology of RES-enabled cellular networks (i.e. physical location of BS and the location of RES) to satisfy coverage in accordance with quality of service (QoS) needs (defined in some specific sense), while significantly reducing capital and operational expenditure (CAPEX and OPEX, respectively). Therefore, inspired by the studies of BSs deployment in traditional cellular networks~\cite{E.Amaldi03,K.Johansson07}, this letter attempts to answer this question by formulating a novel optimization framework for planning the RES-enabled cellular network. As the problem is shown to be NP-hard, we build a heuristic for cellular topology planning consisting of two phases: (i) QoS-aware BS deployment and (ii) energy balancing connection. Numerical results show that the proposed heuristic brings CAPEX and OPEX savings in comparison to traditional network deployment methods.
\vspace{-3.9mm}

\section{System Model}
\label{sec:system_model}

\paragraph*{Network Characterization} We consider an area where a cellular network must be deployed or enhanced/upgraded with new features. As downlink is the bottleneck of a cellular network (since far more traffic is sent over downlink than uplink) it needs to be handled first in network planning. We thus focus on downlink transmission. Then we assume a set of candidate sites for BS installment labeled $\mathcal{N}=\{1,\dotsc,N\}$, with a respective installation cost $c_{n}$, $n\in \mathcal{N}$. Further, we assume a set of test points (TPs), i.e., centroids, where a given amount of traffic or a certain level of QoS service must be guaranteed~\cite[Sec. III-A]{E.Amaldi03}, located at sites $\mathcal{M}=\{1,\dotsc,M\}$. Expressly, TP can be interpreted as a (set of) user equipment in the cellular networks (with certain properties, e.g. minimum required signal to noise ratio, defined individually per each TP).

\IEEEpubidadjcol\paragraph*{Energy Consideration} As BS $n$ is connected to an electric grid consuming power $P_{n}^{(g)}$, $\forall n\in\mathcal{N}$, we aim to minimize the energy consumption at the network planning stage. We thus assume that each BS is equipped with RES, e.g., solar panel, wind turbine. The instantaneous (and aggregated from all energy sources) power delivery capacity of RES for BS $n$, $Z_{n}$, is a stationary stochastic process described by PDF $f_{n}^{(r)}(z_n)$. As a fundamental novelty, for potential energy delivery improvements, we assume that BSs can have \emph{dedicated power line connections for energy balancing}\footnote{Note: this concept has been independently presented in~\cite{chia_wcnc_2012} once our paper was already under review.} from other RESs. On the other hand RESs are not equipped with any battery to store energy (as fixed electric grid connection guarantees a constant energy supply in case of insufficient energy from RES). This assumption results in a lower bound for the system planning, noting however that our model can be extended to consider energy storage. The power transferred from BS $t$ to BS $n$  (through its co-located RES) is denoted as $P_{t,n}$, while the associated (distance-dependent) cost, e.g. installation, maintenance, operation, of connecting BS $t$ and $n$ is denoted as $c_{t,n}$, $t\in\mathcal{N}$.

\paragraph*{Network Planning Constraints} We define the following Boolean variables. The BS deployment indicator $\mathsf{b}_{x}=1$ if a BS is installed on site $x\in\mathcal{N}$, and $\mathsf{b}_{x}=0$ otherwise. The TP assignment indicator $\mathsf{p}_{m,n}=1$ if TP $m\in\mathcal{M}$ is served by BS $n\in\mathcal{N}$, and $\mathsf{p}_{m,n}=0$ otherwise. The inter-RES power line connection indicator for each BS $\mathsf{c}_{t,n}=1$ if BS $t\in\mathcal{N}$, $n\in\mathcal{N}$ are connected by a power line, and $\mathsf{c}_{t,n}=0$ otherwise. We then have
\begin{equation}
\mathsf{p}_{m,n}\leq \mathsf{b}_{n}, \mathsf{c}_{t,n}\leq \mathsf{b}_{n}, \mathsf{c}_{t,n}\leq \mathsf{b}_{t}.
\label{eq:eqn1}
\end{equation}
We further assume the following constraints. Each TP has to be connected to exactly one BS, i.e.,
\begin{equation}
\sum_{n\in \mathcal{N}}\mathsf{p}_{m,n}=1, \forall m\in\mathcal{M}.
\label{eq:eqn2}
\end{equation}
The number of TPs served by each BS is upper bounded, such that
\begin{equation}
\sum_{m\in \mathcal{M}}\mathsf{p}_{m,n}\leq B\mathsf{b}_{n}, \forall n \in \mathcal{N},
\label{eq:eqn3}
\end{equation}
where $B$ is, without a loss of generality, the maximum number of TPs served by each BS, i.e. total radio resources available to each BS. Each BS is assumed to occupy the entire spectrum, i.e., factor-1 reuse strategy is adopted.

\paragraph*{Propagation Environment} We use $\mathbf{H}\triangleq[h_{m,n}]_{M\times N}$ to denote the channel-dependent propagation gain matrix where $h_{m,n}$ is the long-term propagation gain of the link between TP $m$ (averaged over all users in TP $m$) and BS $n$~\cite[Sec. III-A]{E.Amaldi03}. Denoting $\delta_{m}$ as the thermal noise power at TP $m$ and $P_{n}$ as the total transmission power of BS $n$ for each associated TP, we define SINR between TP $m$ and BS $n$ as $\mathsf{SINR}_{m,n}=P_{n}h_{m,n}\left(\sum_{t\neq n}\rho_{t}P_{t}h_{m,t}+\delta_{m}\right)^{-1}$, $\forall m \in \mathcal{M},\ n \in \mathcal{N}$, where $\rho_{t}=\sum_{j\in \mathcal{M}}\mathsf{p}_{j,t}/B$ denotes the ratio of occupied resources to all available resources at BS $t$~\cite[Eq. (10)]{I.Viering09}. In addition, TP $m$ has a minimum QoS requirement defined in SINR terms, $\gamma_{m}$, such that $\mathsf{SINR}_{m,n}\geq \mathsf{p}_{m,n}\gamma_{m}$, $\forall m\in\mathcal{M}, n\in\mathcal{N}$. For the ease of subsequent discussion we transform the non-linear QoS constraint into an equivalent linear formulation through~\cite[Sec. IV-C]{M.Johansson06} as
\begin{align}
\mathsf{SINR}_{m,n}&\geq \mathsf{p}_{m,n}\gamma_{m} \Leftrightarrow\nonumber\\ &M_{m,n}(1-\mathsf{p}_{m,n})+P_{n}h_{m,n}\mathsf{p}_{m,n}\geq\nonumber\\ &\qquad\gamma_{m}\left(B^{-1}\sum_{t\neq n}\sum_{j\in \mathcal{M}}\mathsf{p}_{j,t}P_{t}h_{m,t}+\delta_{m}\right),
\label{eq:eqn4a}
\end{align}
where $M_{m,n}=\gamma_{m}\left(\sum_{t\neq n}P_{t}h_{m,t}+\delta_{m}\right)$ is a sufficiently large constant such that (\ref{eq:eqn4a}) holds for any $\mathsf{p}_{m,n}\in\{0,1\}$, $\forall m \in \mathcal{M},\ n \in \mathcal{N}$.

\paragraph*{Power Outage Constraints} The energy consumption of each BS is limited by the sum of its all energy sources. Thus, an energy outage happens at BS $n$ when its energy supply is insufficient for the energy demand to serve its users. Setting a very small upper bound of the energy outage probability for each BS $n$, $\varphi_{n}$, (to limit energy supply loss due to, e.g., blackout) yields the following constraint
\begin{align}
&\mathsf{b}_{n}\left(\sum_{m\in \mathcal{M}}\mathsf{p}_{m,n}P_{n}+P^{(o)}_{n}-P_{n}^{(g)}+\sum_{t\neq n}P_{n,t}\mathsf{c}_{n,t}-\right.\nonumber\\ &\left.\sum_{t\neq n}(1-\varepsilon_{t,n})P_{t,n}\mathsf{c}_{t,n}-I_{n}^{(r)}(\varphi_{n})\right) \leq 0, \forall n\in \mathcal{N},
\label{eq:eqn5a}
\end{align}
where $\varepsilon_{t,n}\in[0,1]$ is the energy loss factor on the power line between RES $t$ and $n$\footnote{With $\varepsilon_{t,n}$ we guarantee that we do not deploy inter-RES links to transfer negligible amounts of energy between RESs.}, $P^{(o)}_{n}$ is the required static power for operating BS, e.g., cooling, baseband processing~\cite{G.Auer11}, and $I_{n}^{(r)}(\cdot)$ is the inverse of $F_{n}^{(r)}(z_n)=\int_{0}^{z_n}f_{n}^{(r)}(\xi)d\xi$.\vspace{-3.9mm}

\section{Problem Formulation via Optimization}
\label{sec:optimization_framework}

The objective is to select a subset of candidate BSs within $\mathcal{N}$ and to assign TPs  within $\mathcal{M}$ to an available BSs taking into account the QoS requirement, BSs installation, inter-RES connection cost, and the electric grid consumption. Combining the constraints from Section~\ref{sec:system_model} we obtain the following optimization problem
\begin{subequations}
\label{eq:eqn6}
\begin{align}
& \min_{\{\mathsf{p}_{m,n},\mathsf{c}_{t,n},\mathsf{b}_{n},P^{(g)}_{n}\geq0, P_{t,n}\geq 0 \}} \underbrace{\sum_{n\in \mathcal{N}}c_{n}\mathsf{b}_{n}}_{\text{(i)}}+\underbrace{\sum_{t,n \in \mathcal{N}}c_{t,n}\mathsf{c}_{t,n}}_{\text{(ii)}}+\nonumber\\&\qquad\underbrace{\lambda T\left(\sum_{n\in \mathcal{N}} P_{n}^{(g)}\mathsf{b}_{n}+ \sum_{t,n \in \mathcal{N}} \varepsilon_{t,n} P_{t,n}\mathsf{c}_{t,n}\right)}_{\text{(iii)}}
\label{eq:egn6a}
\end{align}
\begin{equation}
\text{subject to  (\ref{eq:eqn1}), (\ref{eq:eqn2}), (\ref{eq:eqn3}), (\ref{eq:eqn4a}), (\ref{eq:eqn5a}).}
\label{eq:6b}
\end{equation}
\end{subequations}
The terms (i) and (ii) correspond to the total installation and the connection cost, respectively. The term (iii) characterizes the cost of consumed power from the electric grid, where $\lambda$ is the energy price expressed in cost unit/kWh, and $T$ is the specified life cycle of the deployed cellular network. Sum of (i)--(iii) terms in (\ref{eq:egn6a}) denotes CAPEX and OPEX. Note that consideration of the random energy arrivals to the harvesters in our planning framework are considered in~(\ref{eq:6b}) through~(\ref{eq:eqn5a}).

\begin{proposition}
Problem (\ref{eq:eqn6}) is NP-hard.
\begin{proof}
(Sketch) Following~\cite{E.Amaldi03}, the problem of (\ref{eq:eqn6}) can be reduced to an uncapacitated facility location problem which is NP-hard~\cite[Eq. (3)--(6) and Sec. IV]{E.Amaldi03} and implies NP-hardness of (\ref{eq:eqn6}).
\end{proof}
\end{proposition}

\vspace{-3.9mm}
\section{Problem Solution via Heuristic}
\label{sec: heuristic_algorithm}

To solve (\ref{eq:eqn6}), we propose a heuristic that decomposes (\ref{eq:eqn6}) into two decoupled subproblems. The first subproblem is the QoS-aware BS deployment (that deploys BSs and connect TPs to BSs according to the energy distribution of BSs, assuming all candidate BSs are disconnected). The second subproblem is the energy balancing (that further reduces the total cost of network operators by balancing the benefits given through energy sharing among deployed BSs and their RESs  and the incurred cost on BSs connection).\vspace{-3.9mm}

\subsection{QoS-Aware BS Deployment}
\label{subsec:QoSaware}

First, we assume that no inter-RES connections are possible, i.e., $\mathsf{c}_{t,n}=P_{t,n}=0$, $\forall n,t \in \mathcal{N}$. Defining $\mathbb{C}_{1}(\mathbf{v},\mathbf{w})=\sum_{n\in \mathcal{N}}c_{n}\mathsf{v}_{n}+\lambda T\sum_{n\in \mathcal{N}} w_{n}\mathsf{v}_{n}$, where $\mathbf{v}=[\mathsf{v}_{n}]_{N\times 1}$, $\mathbf{w}=[w_{n}]_{N\times 1}$,  (\ref{eq:eqn6}) reduces to
\begin{subequations}
\label{eq:eqn7}
\begin{equation}
\min_{\{\mathsf{p}_{m,n}, \mathsf{b}_{n}, P^{(g)}_{n}\geq 0 \}} \mathbb{C}_{1}(\mathbf{b},\mathbf{P})
\end{equation}
\begin{align}
\text{subject to  (\ref{eq:eqn1}), (\ref{eq:eqn2}), (\ref{eq:eqn3}), (\ref{eq:eqn4a}),}\label{eq:eqn7b}
\end{align}
\begin{align}
\!\!\!\mathsf{b}_{n}\left(\sum_{m\in \mathcal{M}}\mathsf{p}_{m,n}P_{n}+P^{(o)}_{n}-P_{n}^{(g)}-I_{n}^{(r)}(\varphi_{n})\right) \leq 0,
\label{eq:eqn7c}
\end{align}
\end{subequations}
where $\mathbf{b}\triangleq[\mathsf{b}_{n}]_{N\times 1}$, $\mathbf{P}\triangleq[P_{n}^{(g)}]_{N\times 1}$.

\begin{proposition}
Suppose an optimal solution to (\ref{eq:eqn7}) exists, denoted as a 3-tuple $\left\{\mathbf{b}^{*}, \mathbf{p}^{*}, \mathbf{P}^{*}\right\}$, where $\mathbf{p}\triangleq[\mathsf{p}_{m,n}]_{M\times N}$ and $*$ symbol denotes optimality. Defining $\Delta_n(\mathbf{u})\triangleq-\sum_{m\in \mathcal{M}}\mathsf{u}_{m,n}P_{n}-P^{(o)}_{n}+I_{n}^{(r)}(\varphi_{n})$, where $\mathbf{u}=[\mathsf{u}_{m,n}]_{M\times N}$, then
\begin{align}
P_{n}^{(g)*}\mathsf{b}_{n}^{*}=\max\left\{-\Delta_n(\mathbf{p}^{*})\mathsf{b}_{n}^{*},0\right\},
\label{eq:eqn8}
\end{align}
where $\Delta_n(\mathbf{p}^{*})$ is the remaining energy from RES applied to transmission and operation of BS.
\begin{proof}
For $\mathsf{b}_{n}^{*}=1$, (\ref{eq:eqn7c}) reduces to $P_{n}^{(g)}\geq -\Delta_n(\mathbf{p})$, which together with $P^{(g)}_{n}\geq 0$ implies that
\begin{equation}
P_{n}^{(g)}\geq \max\left\{-\Delta_n(\mathbf{p}),0\right\}, \ \forall n\in \mathcal{N}.
\label{eq:eqn9a}
\end{equation}

As the objective function of (\ref{eq:eqn7}) is strictly increasing in $P_{n}^{(g)}$, $\forall n\in\mathcal{N}$, when $\mathsf{b}_{n}^{*}=1$, we conclude that (\ref{eq:eqn9a}) will be always tight in the optimal condition, i.e., $P_{n}^{(g)*}= \max\left\{-\Delta_n(\mathbf{p}^{*}),0\right\}$, $\forall n\in \mathcal{N}$. On the other hand, (\ref{eq:eqn8}) holds when $\mathsf{b}_{n}^{*}=0$.
\end{proof}
\end{proposition}
Applying (\ref{eq:eqn8}) to (\ref{eq:eqn7}) yields
\begin{subequations}
\label{eq:eqn9}

\begin{equation}
\min_{\{\mathsf{b}_{n},\mathsf{p}_{m,n}\}} \lambda T \sum_{n \in \mathcal{N}} \max\left\{\frac{c_{n}}{\lambda T }-\Delta_{n}(\mathbf{p}),\frac{c_{n}}{\lambda T }\right\} \mathsf{b}_{n}
\end{equation}
\begin{equation}
\text{subject to (\ref{eq:eqn7b})}.
\end{equation}
\end{subequations}
Notice that solution to (\ref{eq:eqn9}) is equal to the solution to the original problem (\ref{eq:eqn7}). From (\ref{eq:eqn9}) follows that it is beneficial to deploy BSs at sites with positive $\Delta_{n}(\mathbf{p})$, and to connect TPs to the deployed BSs with surplus renewable energy. We devise a heuristic to solve (\ref{eq:eqn9}) (i.e. an algorithm that reaches as close as possible to a final solution) which is presented in Algorithm~\ref{alg:QABD}. Therein, (for ease of exposition) the total radio resources are assumed sufficient for network planning, i.e. $BN\geq M$. Note that Algorithm 1 might not output the assignment for all existing TPs when, e.g., QoS requirements are too strict. In this case, e.g., new positions (or the number) of BS must be considered and Algorithm 1 must run again. The convergence for Algorithm \ref{alg:QABD} can be achieved in at most $N$ outer loops, where the worst case represents reaching $\mathcal{N}=\varnothing$.

\setlength{\textfloatsep}{8pt plus 1.0pt minus 2.0pt}
\begin{algorithm}[t]
\caption{QoS-aware BS deployment}
\label{alg:QABD}
\begin{algorithmic}[1]
\REQUIRE $\forall m, n, t$: $\mathcal{N}$; $\mathcal{M}$; $c_{t,n}$, $P_{n}$, $\delta_m$, $\varphi_n$, $c_n$, $P_{n}^{(o)}$, $B$ (s.t. $BN\geq M$), $\varepsilon_{t,n}$, $\gamma_{m}$; $f_{n}^{(r)}(z_n)$; $\mathbf{H}$
\STATE connect all TPs to closest BSs with free (unassigned) radio resources\label{state:1}
\STATE $\Omega:=\{m|\sum_{n \in\mathcal{N}}\mathsf{SINR}_{m,n}\mathsf{p}_{m,n}<\gamma_{m}, m\in \mathcal{M}\}$, i.e. infeasible TPs
\IF{$\Omega=\varnothing$}
\STATE $F:=1$; $\hat{\mathbf{p}}:=\mathbf{p}$; $\hat{\mathbf{b}}:=\mathbf{b}$; $\hat{\mathbf{P}}:=\mathbf{P}$
\REPEAT \label{state:repeat}
\STATE from BS set with least number of TPs, i.e., $n_{\diamond}\in \arg \min_{n\in\mathcal{N}} \sum_{m\in \mathcal{M}}p_{m,n},$ randomly select one and disconnect its associated TPs, i.e., $\mathcal{M}_{\diamond}$
\FORALL{TP $m\in\mathcal{M}_{\diamond}$}
\STATE find $\mathsf{b}_{n}$ by assigning TP $m$ to the closest BS $n$ (except $n_{\diamond}$) with positive $\Delta_{n}(\mathbf{p})$ without violating (\ref{eq:eqn3})
\IF{such $\mathsf{b}_{n}$ exists}
\STATE $\mathsf{p}_{m,n}=1$, $n\in \mathcal{N}$; $\mathcal{M}_{\diamond}:=\mathcal{M}_{\diamond}\backslash \{\text{TP } m\}$ 
\ELSE 
\STATE $\mathsf{p}_{m,n}=0$, $n\in \mathcal{N}$
\ENDIF
\ENDFOR
\STATE $\Omega:=\{m|\sum_{n \in\mathcal{N}}\mathsf{SINR}_{m,n}\mathsf{p}_{m,n}<\gamma_{m}, m\in \mathcal{M}\}$
\IF{$\Omega=\varnothing$ and $\mathcal{M}_{\diamond}=\varnothing$}
\STATE $\mathcal{N}:=\mathcal{N}\backslash \left\{n_{\diamond}\right\}$; $\hat{\mathbf{p}}:=\mathbf{p}$; $\hat{\mathbf{b}}:=\mathbf{b}$; $\hat{\mathbf{P}}:=\mathbf{P}$
\ENDIF
\UNTIL{$\Omega\neq\varnothing$ or $\mathcal{N}=\varnothing$}\label{state:until}
\ELSE
\STATE $F:=0$
\ENDIF
\ENSURE $F:=0$: no solution found; $F:=1$: $\tilde{\mathbf{p}}:=\hat{\mathbf{p}},\tilde{\mathbf{b}}:=\hat{\mathbf{b}},\tilde{\mathbf{P}}:=\hat{\mathbf{P}}$, i.e. optimized BS configuration \label{state:2}
\end{algorithmic}
\end{algorithm}

\vspace{-3.9mm}
\subsection{Energy Balancing Inter-RES Connection}
\label{subsec:energybalancing}

Let $\mathcal{\tilde{N}}$ denote the set of deployed BSs based on $\tilde{\mathbf{b}},\tilde{\mathbf{p}},\tilde{\mathbf{P}}$ (i.e. the output of Algorithm~\ref{alg:QABD}). Defining $\mathbb{C}_{2}(\mathbf{x},\mathbf{y},\mathbf{z})\triangleq\lambda T\left\{\sum_{t,n \in \mathcal{\tilde{N}}}\left(\varepsilon_{t,n} z_{t,n}\mathsf{c}_{t,n}\!+\!\frac{x_{t,n}}{\lambda T}\mathsf{c}_{t,n}\right)\!+\!\sum_{n\in \mathcal{\tilde{N}}} \left(y_{n}\!\!+\!\frac{c_{n}}{\lambda T}\right)\right\}$, where $\mathbf{x}=[x_{t,n}]_{N\times N}$, $\mathbf{y}=[y_{n}]_{N\times 1}$, and $\mathbf{z}=[z_{t,n}]_{N\times N}$. After the QoS-aware BS deployment shown in Section~\ref{subsec:QoSaware}, (\ref{eq:eqn6}) further reduces to
\begin{subequations}
\label{eq:eqn10}
\begin{align}
\min_{\{\mathsf{c}_{t,n},P^{(g)}_{n}\geq0,P_{t,n}\geq0 \}} & \mathbb{C}_{2}(\mathbf{C},\mathbf{P},\mathbf{R})
\end{align}
\begin{equation}
\text{subject to (\ref{eq:eqn5a}) with } \mathsf{p}_{m,n} \text{ from } \tilde{p} \text{, replacing } \mathcal{N} \text{ by } \mathcal{\tilde{N}},
\end{equation}
\end{subequations}
where $\mathbf{C}\triangleq[\mathsf{c}_{t,n}]_{N\times N}$ and $\mathbf{R}\triangleq[P_{t,n}]_{N\times N}$.

We propose a second heuristic (algorithmic solution) given in Algorithm~\ref{alg:EBC} to reach as close as possible to a solution of (\ref{eq:eqn10}). The heuristic starts with a fully connected RESs topology, with a set of inter-RES connections denoted as $\mathcal{E}\triangleq\{e_{t,n}\}_{t,n \in \mathcal{\tilde{N}}}$. Within each iteration of the heuristic, the inter-RES power lines transferring the least amount of energy are removed from $\mathcal{E}$ until no cost saving can be achieved. Notice that $\{\tilde{\mathbf{C}}=[1]_{\forall t,n};\tilde{\mathbf{P}};\tilde{\mathbf{R}}=[0]_{\forall t,n}\}$ is a feasible solution to (\ref{eq:eqn10}), thus the feasibility of Step \ref{state:lp} in Algorithm \ref{alg:EBC} always holds. In addition, termination criteria for the loop (Step \ref{state2:repeat}--\ref{state2:until}) in Algorithm \ref{alg:EBC} are achievable within $|\mathcal{E}|$ loops.

\begin{algorithm}[t]
\caption{Energy Balancing inter-RES Connection Deployment}
\label{alg:EBC}
\begin{algorithmic}[1]
\REQUIRE complete set of inter-RES connections $\mathcal{E}$; $\forall t,n$: $\varepsilon_{t,n}$, $c_{n}$; $\tilde{\mathbf{p}},\tilde{\mathbf{b}},\tilde{\mathbf{P}}$ from Algorithm~\ref{alg:QABD};  $\mathbb{S}\triangleq\infty$
\REPEAT \label{state2:repeat}
\STATE find $\tilde{\mathbf{C}}$, $\tilde{\mathbf{P}}$, $\tilde{\mathbf{R}}$, i.e. the solution of (\ref{eq:eqn10}), with $\tilde{\mathbf{p}},\tilde{\mathbf{b}}$, $\mathsf{c}_{t,n}=1$ $\forall {e}_{t,n}\in \mathcal{E}$ \label{state:lp}
\STATE target zero and the smallest positive $P_{t,n}$ in $\tilde{\mathbf{R}}$, then remove corresponding connections from $\mathcal{E}$, i.e., $\mathcal{E}:=\mathcal{E}\backslash\{e_{t,n}\}$
\STATE $\mathbb{S}:=\min\left\{\mathbb{S},\mathbb{C}_{2}(\tilde{\mathbf{C}},\tilde{\mathbf{P}},\tilde{\mathbf{R}})\right\}$
\UNTIL{$\mathbb{C}_{2}(\tilde{\mathbf{C}},\tilde{\mathbf{P}},\tilde{\mathbf{R}})>\mathbb{S}$ or $\mathcal{E}=\varnothing$}\label{state2:until}
\IF{$\mathbb{C}_{1}(\tilde{\mathbf{b}},\tilde{\mathbf{P}})<\mathbb{S}$}
\STATE $\mathcal{E}:=\varnothing$
\ENDIF
\ENSURE optimized inter-RES connections according to $\mathcal{E}$
\end{algorithmic}
\end{algorithm}
\vspace{-3.9mm}

\subsection{Heuristics Complexity Analysis}
\label{sec:complexityanalysis}

The complexity of Algorithm \ref{alg:QABD} is determined by the double-nested loop (Step~\ref{state:repeat}--\ref{state:until}) whose complexity is $O(MN^{2})=O(N^{3})$ since $M=O(N)$, i.e., the number of TPs is limited by $M\leq BN$. The complexity of Algorithm~\ref{alg:EBC} is governed by the time needed to solve a Linear Program (LP) (Step~\ref{state:lp}). As each LP can be solved in polynomial time~\cite[Theorem 8.5]{Papadimitriou82} its complexity is $O(X^t)$, where $X$ is the number of LP variables and $t$ characterizes the running time of the LP solver. Thus the complexity of Algorithm~\ref{alg:EBC} is $O(X)O(X^{t})$ for $X=N^{2}$. In turn, the complexity of the complete heuristic (Algorithm~\ref{alg:QABD} and Algorithm~\ref{alg:EBC}) is $O(N^{2t+2})$.\vspace{-3.9mm}

\section{Numerical Results}
\label{sec:numerical_results}

As an illustrative simple example we consider a 3\,km$\times$3\,km area where $m$ TPs are arbitrarily located and $n$ BS candidate sites (forming a grid) are identified to provide services for TPs. For a cost unit of \euro~we assume all RES-enabled BSs have fixed and equal installation cost, $c_n$~\cite[Sec. II-B]{brevis_vtc_2011} and fixed transmission power~\cite[Sec. II-C]{M.Johansson06}. Furthermore, for a fair comparison we assume that all traditional (no-RES enabled) BSs have an installation cost of $\bar{c}_n$ and find optimal cellular network structure using Algorithm~\ref{alg:QABD} considering no RES and no inter-RES connections. Assuming distance-dependent pathloss-only scenario, the elements of $\mathbf{H}$ are given as $h_{m,n}=10^{\frac{-\{L_{A}+L_{B}\log_{10}(d_{m,n}/\text{km})\}}{10}}$~\cite[Eqs. (1), (7)]{I.Viering09}, where $d_{m,n}$ denotes the distance between TP $m$ and BS $n$ and $L_A$ and $L_B$ are the empirical constants provided by~\cite{3GPP10}.

As the distribution of harvested power at location $n$ is scenario-dependent, for simplicity we assume it being uniformly distributed, i.e., $f_{r}^{(n)}(z_n)=\frac{1}{b_n-a_n}$, $\forall z_n\in[a_n,b_n]$, $n\in \mathcal{N}$, with $a_n$ and $b_n$ being the minimum and maximum harvested power at location $n$, respectively (note that we can use any continuous energy distribution in our model). All deployment parameters are summarized in Table~\ref{tab:specification}, representing values describing typical network scenario, following, e.g.~\cite{E.Amaldi03,K.Johansson07,3GPP10}. Numerical results are generated based on the method of batch means with 100 simulation runs for the confidence level of 95\%. Due to space constraints we focus on two most representative cases.\vspace{-3.9mm}

\begin{table}
\caption{Parameter Values Used in the Numerical Evaluation}
\centering
\scriptsize
\begin{tabular}{@{\hspace{0.05cm}}c@{\hspace{0.05cm}}|@{\hspace{0.05cm}}c@{\hspace{0.05cm}}||@{\hspace{0.05cm}}c@{\hspace{0.05cm}}|@{\hspace{0.05cm}}c@{\hspace{0.05cm}}}
\hline
No. candidate sites & $N=9$ & Network life cycle & $T=[6,20]$\,years\\
\hline
No. TPs & $M=20$ & Electricity price & $\lambda=0.3$\,\euro/kWh \\
\hline
No RES-based BS depl. & $\bar{c}_{n}=55$\,k\euro & RES-based BS depl. & $c_{n}=60$\,k\euro \\
\hline
Inter-RES conns. cost & $c_{t,n}=10$\,\euro/m & BSs oper. power & $P_{n}^{(o)}=19$\,W \\
\hline
Transmission power & $P_{n}=20$\,W & No. res. blocks & $B=\{6,12\}$\ \\
\hline
Thermal noise power & $\delta_{m}=-114$\,dBm & SINR requirement & $\gamma_{m}=0$\,dB \\
\hline
Propagation coeff. A & $L_{A}=148.1$\,dB & Min harv. power & $a_n= [0,100]$\,W \\
\hline
Propagation coeff. B & $L_{B}=37.6$\,dB & Max harv. power & $b_n= [100,200]$\,W \\
\hline
Power outage prob. & $\varphi_{n}=5\%$ & Energy loss & $\varepsilon_{t,n}=1\%$ \\
\hline
\end{tabular}
\label{tab:specification}
\vspace{1mm}
\\ coeff.--coefficient; conns.--connections; depl.--deployment; harv.--harvested; no.--number; oper.--operational; prob.--probability; res.--resource; reported values are equal $\forall m,n,t$
\end{table}

\subsection{CAPEX and OPEX versus Network Life Cycle}
\label{sec:opex_lifetime}

\begin{figure}
\centering
\subfigure[\scriptsize $\times$: No REs (optimal depl. via FICO\textregistered Xpress); $\circ$: No inter-RES conns. (Alg.~\ref{alg:QABD}); $*$: No RES (Alg.~\ref{alg:QABD}); $\diamond$: Inter-RES conns. (Alg~\ref{alg:QABD}.+Alg.~\ref{alg:EBC})]{\includegraphics[height=0.35\columnwidth]{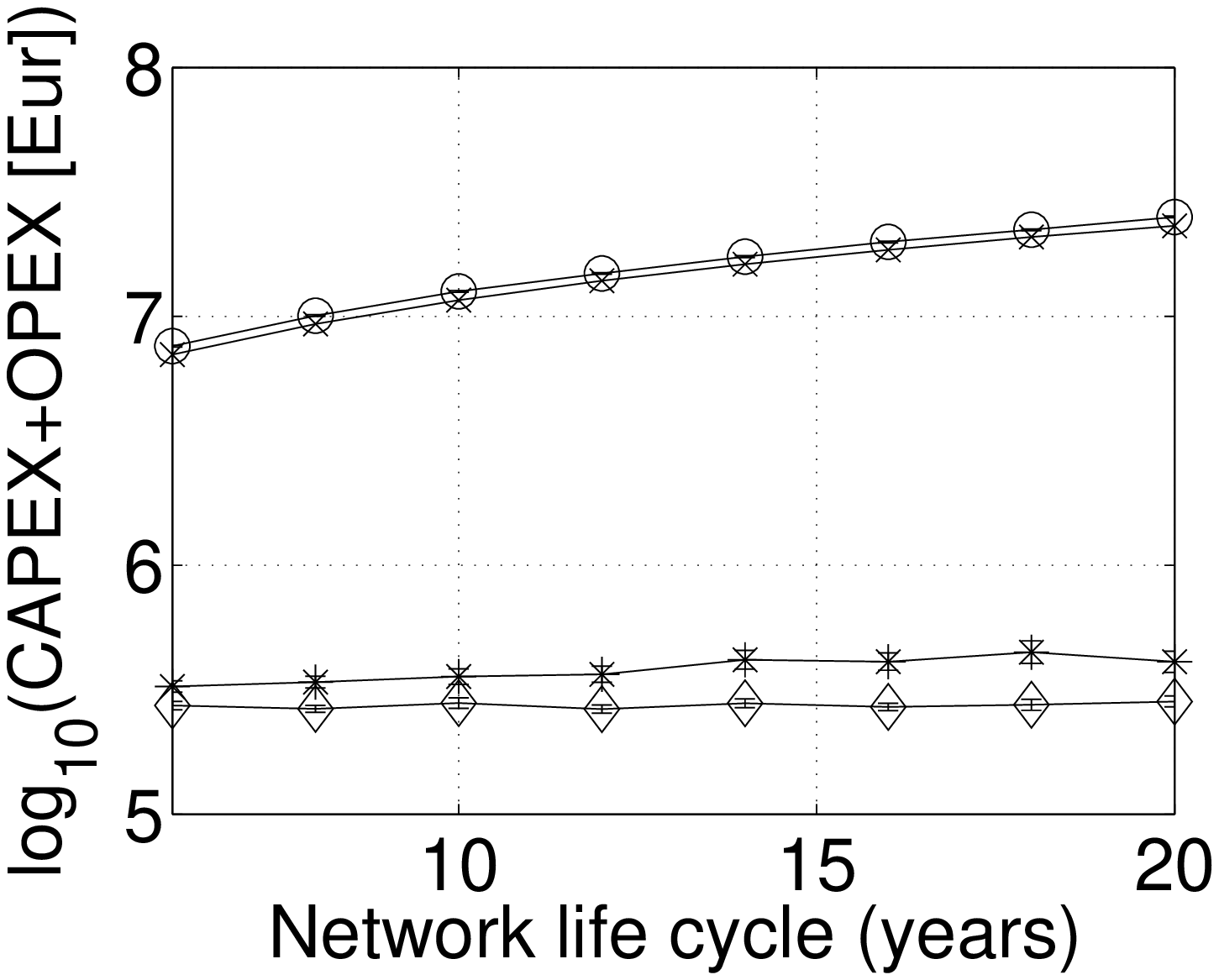}\label{fig:lifetime}}
\subfigure[\scriptsize $\times$: No inter-RES conns. ($B=6$); $\circ$: Inter-RES conns ($B=6$); $*$: No inter-RES conns. ($B=12$); $\diamond$: Inter-RES conns. ($B=12$)]{\includegraphics[height=0.35\columnwidth]{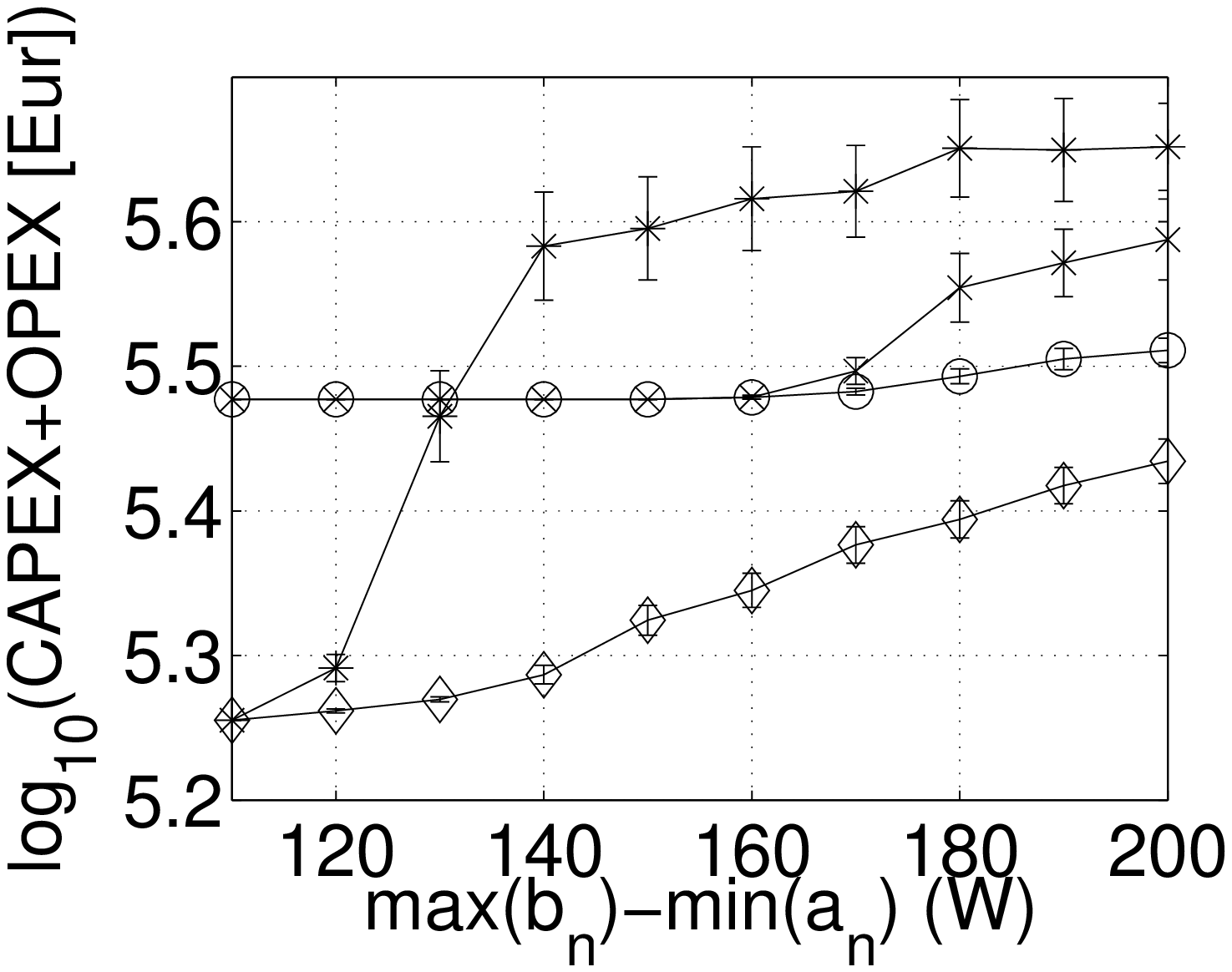}\label{fig:variance}}
\caption{CAPEX and OPEX as a function of: (a) network life cycle $T$, for randomly chosen $a_n$, $b_n$ from Table~\ref{tab:specification} for each location $n$ and $B=12$; (b) harvested energy spread $\max b_n-\min a_n$ for $T=10$ years, where for each simulation point (from left to right) $k=0,\ldots,9: E[z_n]=145-5k$\,(W); whiskers denote 95\% confidence interval; alg.--algorithm; depl.--deployment; conns.--connections.}
\label{fig:results}
\end{figure}

The result is presented in Fig.~\ref{fig:lifetime}. For each simulation run new location points for BS and TPs (with the respective $\mathbf{H}$) have been randomly generated. We immediately observe that the introduction of RESs into network planning brings cost saving to the operator, compared to the traditional deployment structure. This benefit is slightly boosted by the energy balancing among RESs, as for the considered network configuration, per BS, most of the energy is provided by the RES alone. Notice that CAPEX and OPEX with inter-RESs connections does not vary with the network life cycle, as the whole network is basically self-powered (without grid power). Comparing the effectiveness of the developed algorithmic solution (considering the lack of RESs) with the solution to (\ref{eq:eqn6}), computed using FICO$^\text{\textregistered}$ Xpress Optimization Suite version 1.23.00, for 10 independent realizations of $\mathbf{H}$, we conclude that our heuristic is close to the optimal deployment.\vspace{-3.9mm}

\subsection{CAPEX and OPEX versus Harvested Energy Spread}
\label{sec:opex_distribution}

The result is given in Fig.~\ref{fig:variance} for one randomly generated set of TP locations, with the respectively generated $\mathbf{H}$ (one time only). CAPEX and OPEX roughly increases as the difference between $\min a_n$ and $\max b_n$ becomes larger, irrespective of $B$. This is because of strong randomness in available energy, which in turn incurs extra cost on electricity supply (no inter-RES connections) or power lines (inter-RES connections). In addition, we also observe that deploying inter-RES connections with large $B$ is the most cost-effective option for network operators. This is because BSs with large $B$ normally allow the small number of deployed BSs thus saving the deployment cost, while also lead to large energy deficiency at some 'crowed' BSs. This is especially visible with high variance of harvested power, which increases a benefit of RES use by deploying inter-RES connections.\vspace{-3.9mm}

\section{Conclusions}
\label{sec:conclusion}

In this letter we have developed a novel cellular network planning framework considering the use of renewable energy sources and energy balancing. For the posed problem we have developed a novel heuristic. Our numerical results demonstrate CAPEX and OPEX savings in comparison to traditional deployment strategies.

\end{document}